\documentclass[mathleft
]{an}
\usepackage{aas_macros}
\usepackage[sort,authoryear]{natbib}
\usepackage{graphicx}
\usepackage{times}
\bibpunct{(}{)}{;}{a}{}{,}
\begin{document}

\Pagespan{1099}{1102}
\Yearpublication{2012}%
\Yearsubmission{2012}%
\Month{12}%
\Volume{333}%
\Issue{10}%
\DOI{10.1002/asna.201211793}%

\title{EXOTIME: searching for planets and measuring Pdot in
  sdB pulsators}

\author{R. Lutz\inst{1}
  \and
  S. Schuh\inst{1}\fnmsep\thanks{Corresponding author:
    \email{schuh@astro.physik.uni-goettingen.de}\newline}
  \and
  R. Silvotti\inst{2}
}
\titlerunning{EXOTIME: searching for planets and measuring Pdot in
  sdB pulsators}
\authorrunning{R. Lutz, S. Schuh and R. Silvotti}
\institute{
  Georg-August-Universit\"at G\"ottingen, Institut f\"ur Astrophysik,
  Friedrich-Hund-Platz~1, 37077 G\"ottingen, Germany
  \and
  INAF-Osservatorio Astrofisico di Torino, Strada
  dell'Osservatorio 20, 10025, Pino Torinese, Italy 
}

\received{2012 September 5}
\accepted{2012 October 22}
\publonline{2012 December 3}


\abstract{%
We review the status of the EXOTIME project (EXOplanet
search with the TIming MEthod). The two main goals of
EXOTIME are to search for sub-stellar companions to sdB
stars in wide orbits, and to measure the secular variation
of the pulsation periods, which are related to the
evolutionary change of the stellar structure. Now, after
four years of dense monitoring, we start to see some results and
present the brown dwarf and exoplanet candidates V1636\,Ori\,b 
and DW\,Lyn\,b.}
\maketitle
\vspace*{-7mm}
\section{Introduction}
\newcommand{\gsim}{\raisebox{-0.6ex}{$\stackrel{{\displaystyle>}}{\sim}$}}
Subdwarf B (sdB) stars are extreme horizontal branch (EHB) core He-burning
stars, and are characterized by an extremely thin H envelope (see
\citealt{2009ARA&A..47..211H} for a comprehensive review).  They are the
product of a small fraction of red giants (RGs), of the order of 1\%, that
loose almost all their envelope near the tip of the RG branch.
While half of sdB stars are in binary systems and their formation can be
explained in terms of binary evolution
\citep{2002MNRAS.336..449H,2003MNRAS.341..669H}, it is more difficult to
explain the formation of the other half of apparently single sdBs.  A planet
with a sufficient mass (\gsim 10 M$_{\rm Jup}$) and sufficiently close to the
star may have enough energy and angular momentum to remove the RG envelope and
trigger the formation of an sdB star \citep{1998AJ....116.1308S}.  The opposite 
effects of stellar mass loss (which causes an outward drift of the planet) and 
tidal forces (which causes an inward drift), may create a gap in a certain range 
of orbital distances \citep[e.g.][]{2010MNRAS.408..631N}.
\par
Presently, we know a handful of sdB stars (mainly single)
that show some evidence of sub-stellar companions. They
belong to three different groups, 
well distinct in terms of orbital distance and planetary mass. In order of
increasing orbital distance we find:
1) at least five close brown dwarf (BD) candidates with orbital periods of the order 
of 1 day or less \citep{2012ASPC..452..153G}.
2) Two Earth-mass planet candidates around the sdB pulsator KIC~05807616,
with orbital periods of 5.8 and 8.2 hours \citep{2011Natur.480..496C}.
3) Six planet/BD candidates with minimum masses between $\sim$2 and $\sim$40 M$_{\rm Jup}$
in large orbits with orbital periods between 3.2 and $\sim$16 yrs
(\citealt{2007Natur.449..189S};
\citealt{2009AJ....137.3181L}, orbital solutions
significantly revised by \citealt{2012A&A...543A.138B};
\citealt{2009ApJ...695L.163Q}, orbital solution improved by \citealt{2012A&A...540A...8B};
\citealt{2012ApJ...745L..23Q};
\citealt{2012A&A...540A...8B}).
\par
\begin{figure}[t]
\includegraphics[height=0.44\textwidth]{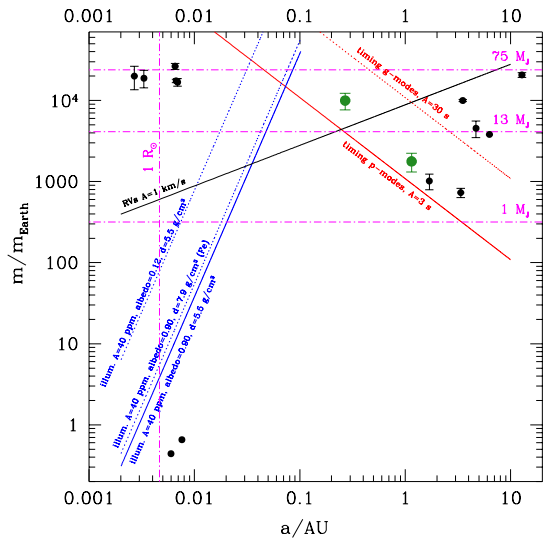}
\caption{Substellar companions around subdwarf B stars (black dots) and comparison
of the detection methods mentioned in the text (RV, illumination and timing). For most of 
these systems the inclination is not known and we report the minimum mass. Horizontal
lines show 1 Jupiter mass, 13 Jupiter masses (dividing the exoplanet and brown dwarf regimes)
and 75 Jupiter masses (dividing the substellar and stellar regimes). The big dots are the two
candidates presented in this paper.\vspace*{-5mm}
} 
\label{detectionmethods}
\end{figure}

These three groups correspond to three different detection methods: radial
velocities (RVs), illumination effects, and timing (using either the stellar
pulsation or the eclipses as a clock). Figure \ref{detectionmethods} 
compares the sensitivities of these different detection methods. 
\par
The EXOTIME project is primarily oriented to increase the statistics of sdB
planets/BDs in large orbits (third group) and particularly focuses on \emph{single}
stars which are not part of a binary system.
The method is based on getting a dense coverage of time-series photometric
data and searching for periodic (pulsation) phase variations due to the star
wobble caused by a planet.
It is the same method that has been used for V391~Peg \citep{2007Natur.449..189S}.
Obviously, the large data sets that we are collecting for our five EXOTIME
targets allow to study also other pulsational properties of the star, in
particular the stability of the pulsation periods (measuring their secular
variation $\dot{P}$) and the amplitude variations, and to improve the star
characterization through asteroseismology.
\par
More details on the EXOTIME project can be found in
\citet{2010Ap&SS.329..231S} and on the EXOTIME web site at 
\textit{http://www.na.astro.it/$\sim$silvotti/exotime/}.
In this paper, we summarize the results obtained for the two new
well-observed targets V1636\,Ori and DW\,Lyn.
\section{Methods}
\subsection{Observations}
In order to measure secular period variations and to search for
companions in wide orbits around sdB stars, we need to
gather data spread over a very long time-baseline. Within
EXOTIME, this is realized via a regular long-term monitoring
of our targets with many small- to medium-class
optical telescopes by means of time-resolved ground-based
photometry. On average, data for each target are taken once
a month on several consecutive nights, to achieve a
sufficient frequency resolution within these observing
blocks. Each of these blocks of roughly 3-5 nights finally
yields one data point for the subsequent O--C analysis,
which will be addressed in the next subsection.
\par
The photometric data for V1636\,Ori were taken with four
different observatories yielding a total amount of 92.7 hours 
in the period from August 2008 to November 2010. DW\,Lyn was
observed with ten different telescopes during November 2007
and November 2010 and the data sum up to 857.5 hours of
photometry.
\subsection{Data analysis}
An O--C diagram is a way to investigate the temporal
behavior of periodic events. In our case, these periodic
events are the intensity variations due to the pulsations of
the sdB stars. In contrast to the solar-like oscillations
with random phases and short lifetimes, the pulsations of
hot subdwarf stars are very stable on long timescales. This
is a major prerequisite to conduct the phase analysis.
\par
In general there may appear different components in an O--C
diagram. Depending on their physical nature, these can be
i) linear if the period is constant (with or without a slope
depending if the guess period is wrong or right),
ii) parabolic if the
period is changing linearly with time, or iii) sinusoidal if
the period increases and decreases periodically. The
secular period change due to the momentary evolution of the star 
results in a parabolic O--C component. The same
holds for the signature due to proper motion. The signature
that a possible companion would induce is a sinusoidal O--C
component due to the companion-induced wobble of the star
around the common barycenter. Another effect resulting in a
sinusoidal O--C component is the beating of close unresolved
frequencies. All these possible components are thoroughly
investigated in \citet{2011PhDT........85L}. In order to 
finally obtain the O--C diagrams shown in
Figs.~\ref{hs0444f1} and \ref{hs0702f1}, we i) subtracted the
parabolic evolutionary component, ii) neglected the
contribution due to proper motion since it is estimated to
be two to three orders of magnitude weaker than the
evolutionary component, iii) subtracted beating signals after
confirming their presence with an independent method and iv)
finally investigated these residual O--C diagrams for
periodic signatures.
\par
We refer to chapter 5 and chapter 6.3 in
\citet{2011PhDT........85L} for a detailed description of
the construction and analysis of the O--C diagrams shown here and
especially for the treatment of the various different O--C 
components that have been mentioned above.
\section{Status on individual targets}
We present first results for the two targets
V1636\,Ori and DW\,Lyn. A previous status report is given in
\citet{2011AIPC.1331..155L}.
Concerning the other EXOTIME targets, new data on
V391\,Peg and QQ\,Vir are under analysis.
\subsection{V541\,Hya (EC09582$-$1137)}
This is a southern object and the least well observed target in the
EXOTIME project (only approx.\ $15$ hours of observations). The
low declination of this star and the low amplitudes result in limited 
observabilities and a restriction to larger mirror-sizes (above 2\,m). 
With the current low coverage, it would probably
take a few more years to achieve reliable O--C diagrams.
\subsection{QQ\,Vir (PG1325+101)}
A status report on QQ\,Vir is given in \citet{2010arXiv1012.0747B}.
Encouraged by the short-term phase stability of this target, QQ\,Vir 
is now the highest priority target of EXOTIME. The data amount for 
this target, gathered between 2008 and 2011, sum up to roughly $124$ 
hours of photometry and are currently under analysis. 
\subsection{V391\,Peg (HS2201+2610)}
Since the publication of a planetary companion candidate
with a minimum mass of $3.2$ Jupiter masses at an orbital
distance of about $1.7$ AU from its host star
\citep{2007Natur.449..189S}, we have continued to monitor
this system in order to verify and consolidate our
results. The new data are under analysis and
an identification of the two main p-modes, based on ULTRACAM 
multicolor photometry, was presented by \citet{2010AN....331.1034S}. 
\begin{figure}[t]
\includegraphics[height=0.44\textwidth,angle=90]{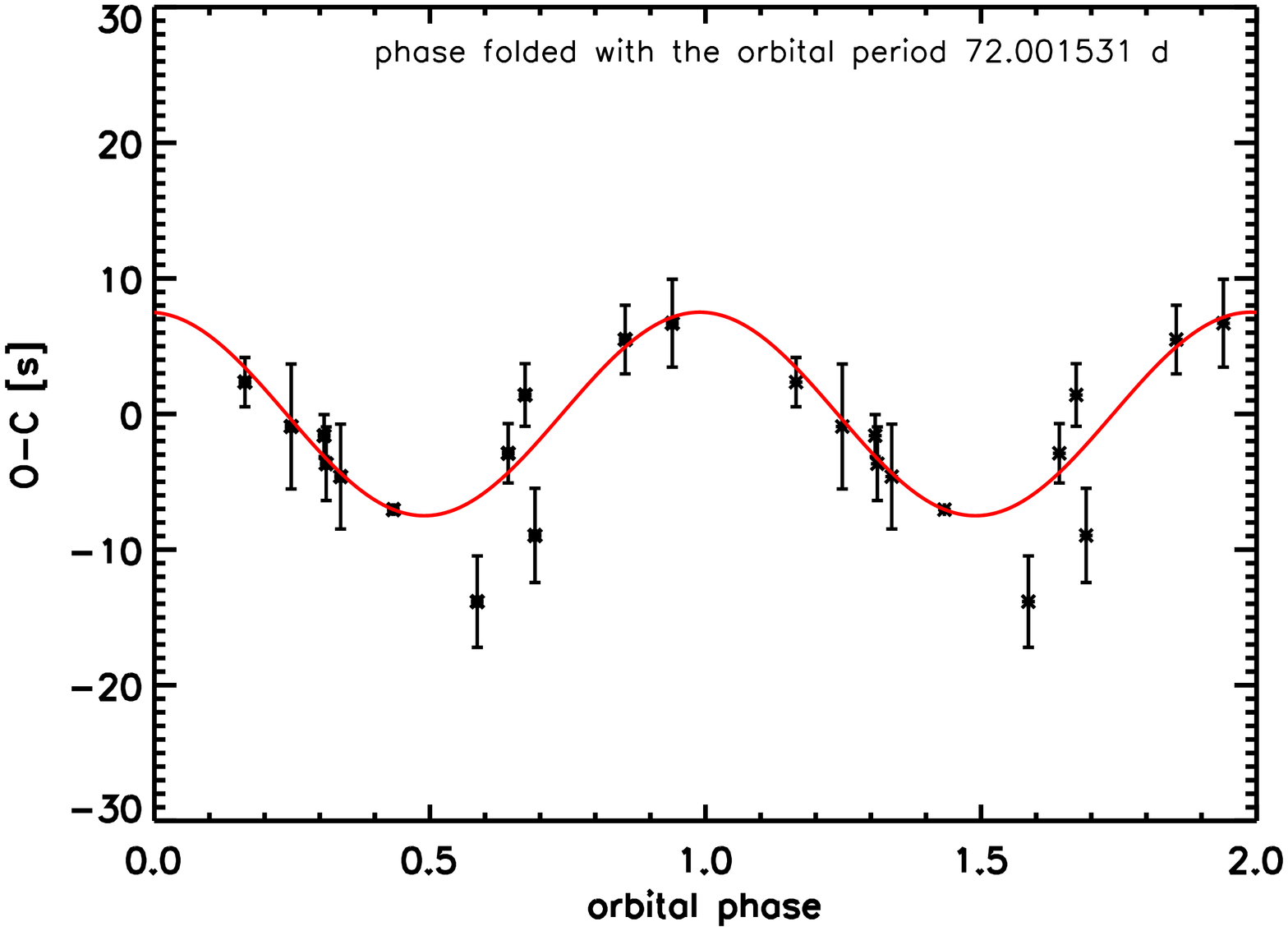}
\includegraphics[height=0.44\textwidth,angle=90]{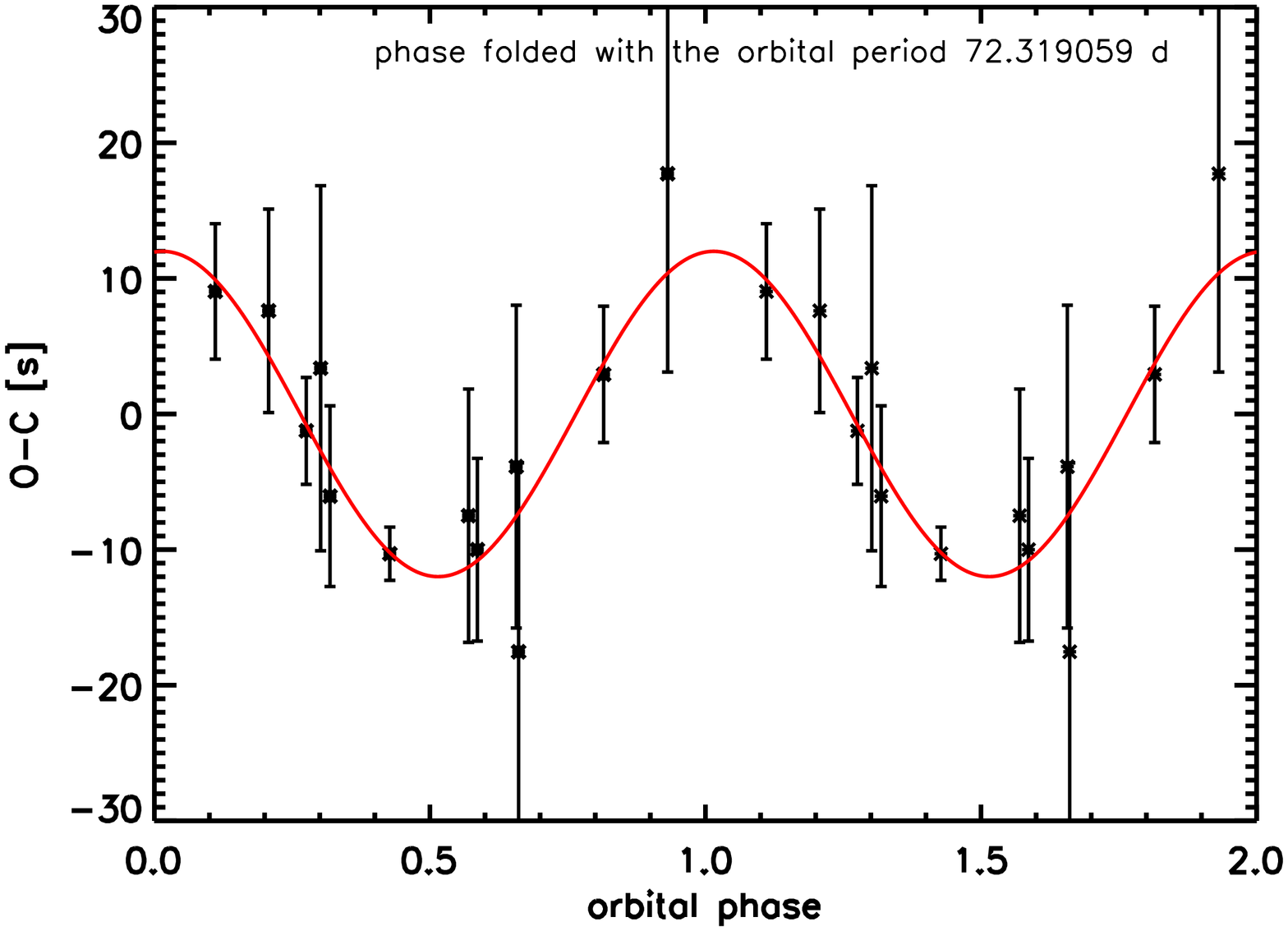}
\caption{Phase-folded O--C diagrams for V1636\,Ori based on the independent frequencies
  $f_1$ (top) and $f_2$ (bottom). Evolutionary and beating signals are already subtracted. 
  The data are folded with an orbital period as given in the plots. Data points are
  duplicated on the phase-axis for plotting purposes.\vspace*{-5mm}} 
\label{hs0444f1}
\end{figure}
\subsection{V1636\,Ori (HS0444+0458)}
\subsubsection{New results on evolutionary time scale}
From the parabolic O--C component, we derive a $\dot{P_1}$ of $(9.300\pm
0.0023)\cdot 10^{-12}$ d/d for \smash{$f_1$} (631.73495 1/d) and $\dot{P_2}$
of $(3.010\pm 0.0010)\cdot 10^{-12}$ d/d for \smash{$f_2$} (509.97773 1/d).
The proper motion contribution to the parabolic component is estimated to be
at least two orders of magnitude smaller than the effect of period evolution
\citep{2011PhDT........85L}. Hence, the proper motion is neglected in our
analysis. These $\dot{P}$ values translate to evolutionary timescales of
\smash{$T_{\textrm{\footnotesize evol},f_1}=0.466\pm 0.002$} Myr and
\smash{$T_{\textrm{\footnotesize evol},f_2}=1.785\pm 0.006$ Myr} for \smash{$f_1$} and
\smash{$f_2$}, respectively.  From the positive sign of $\dot{P}$, we can tell
that V1636\,Ori is in an expansion phase. Additionally, the radial
expansion timescale $\dot{R}$ ($\dot{P}/P \approx 3/2 \dot{R}/R$) suggests
that this star has probably already undergone core helium exhaustion
\cite[compare e.g.\ to Fig.~3 in][]{2010AN....331.1020K}.
\subsubsection{New results on companion search}
After subtracting the evolutionary component,
we find sinusoidal signatures in the O--C diagram of both frequencies
\smash{$f_1$} and \smash{$f_2$}. The most plausible explanation for this
sinusoidal behavior is an orbital origin, hence the presence of a companion
candidate around V1636\,Ori. Figure~\ref{hs0444f1} shows the O--C diagrams for
\smash{$f_1$} and \smash{$f_2$}, phase-folded with an orbital period close to 72 days. 
Within the error bars, the same signature consistent in amplitude, phase and period is
independently also seen in the O--C diagram for \smash{$f_2$}. From the
O--C amplitude and period we can derive the minimum companion mass and orbital
separation. Assuming a canonical sdB mass of $0.47 M_{\odot}$ and weighting
over the independent frequencies \smash{$f_1$} and \smash{$f_2$}, we find a
minimum mass $m\cdot \sin{i}$ of $31.41\pm 7.25$ Jupiter masses for this
candidate, V1636\,Ori\,b, which would place it to the brown dwarf regime
(see Fig.~\ref{inclinations} for the effect of the unknown inclination on the true
mass). The orbital separation is derived to be $0.269\pm 0.001$ AU.
\begin{figure}[t]
\includegraphics[height=0.44\textwidth,angle=90]{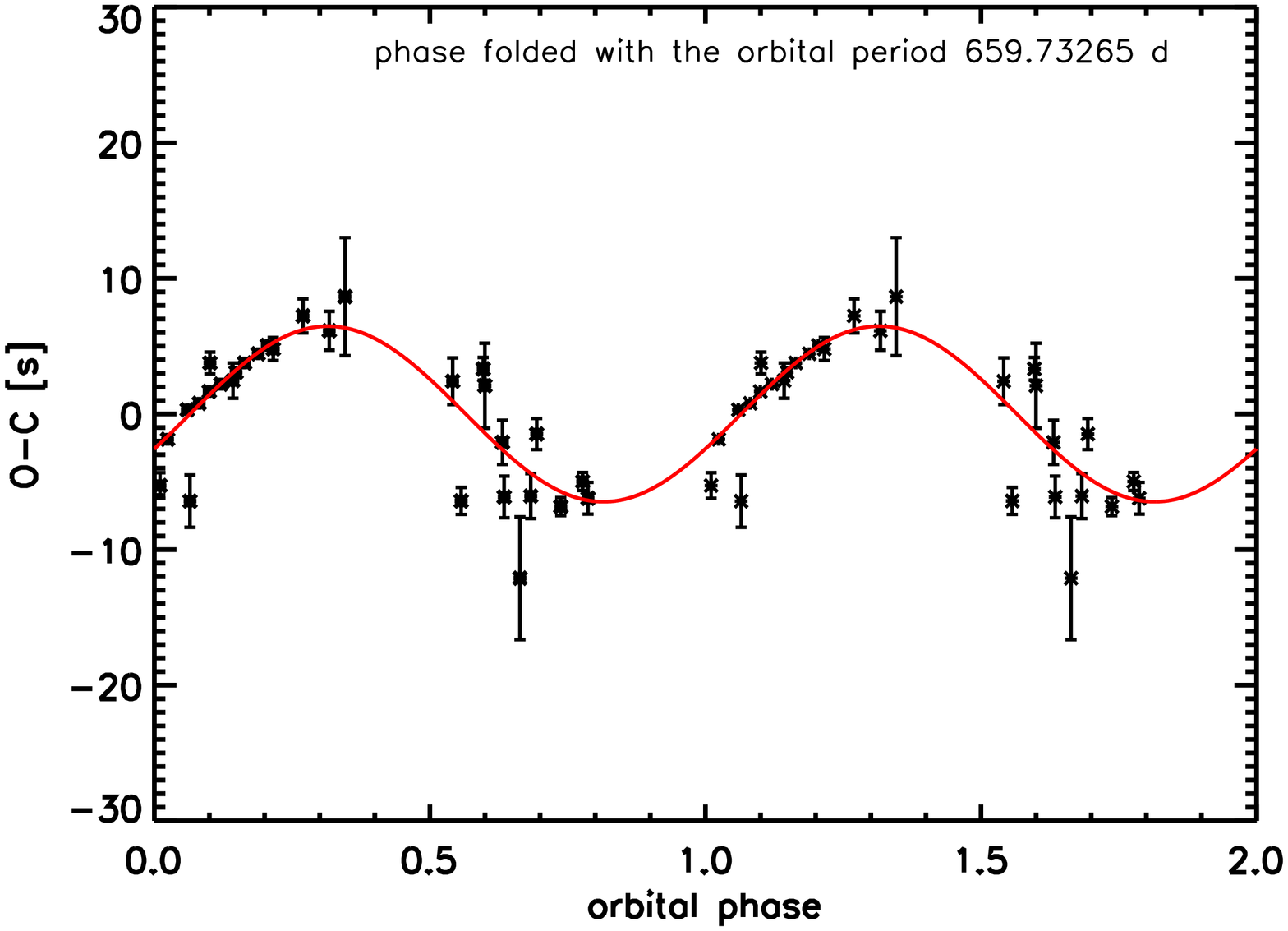}
\includegraphics[height=0.44\textwidth,angle=90]{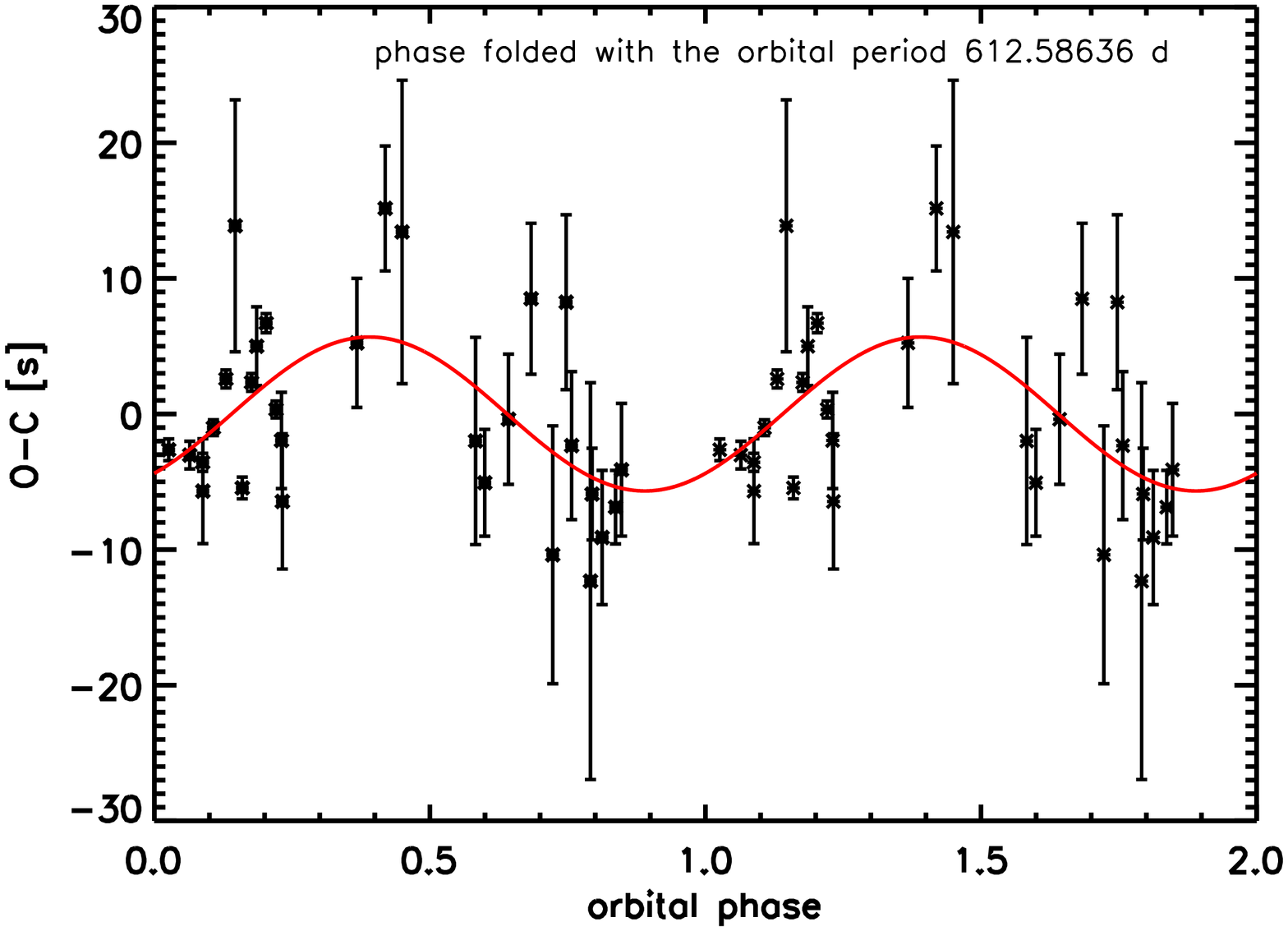}
\caption{Phase-folded O--C diagrams for DW\,Lyn based on the independent frequencies
  $f_1$ (top) and $f_2$ (bottom). Evolutionary and beating signals are already subtracted. 
  The data are folded with an orbital period as given in the plots. Data points are
  duplicated on the phase-axis for plotting purposes.\vspace*{-5mm}} 
\label{hs0702f1}
\end{figure}
\subsection{DW\,Lyn (HS0702+6043)}
\subsubsection{New results on evolutionary time scale}
For DW\,Lyn we derive a $\dot{P_1}$ of $(2.871\pm 0.0025)\cdot 10^{-13}$
d/d for \smash{$f_1$} (237.94106 1/d) and $\dot{P_2}$ of $(5.578\pm
0.006)\cdot 10^{-12}$ d/d for \smash{$f_2$} (225.15882 1/d).  The proper motion
contribution to the parabolic component is neglected here with the same
argument as for V1636\,Ori. The $\dot{P}$ values translate to evolutionary
timescales of \smash{$T_{\textrm{\footnotesize evol},f_1}=40.110\pm 0.350$} Myr
and \smash{$T_{\textrm{\footnotesize evol},f_2}=2.181\pm 0.003$} Myr for
\smash{$f_1$} and \smash{$f_2$}, respectively.  From the positive sign of
$\dot{P}$, we can tell that DW\,Lyn is in an expansion phase. Likewise
for V1636\,Ori, the radial expansion timescale $\dot{R}$ for DW\,Lyn
suggests that this star has probably also undergone core helium exhaustion.
\subsubsection{New results on companion search}
\begin{figure}
\includegraphics[height=0.45\textwidth,angle=90]{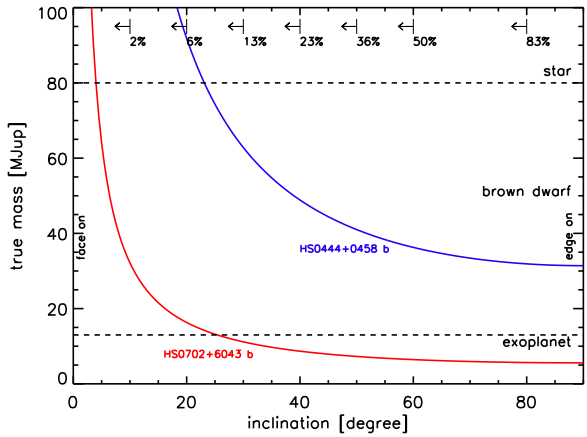}
\caption{True companion masses as a function of the unknown inclinations for
V1636\,Ori\,b (HS\,0444+0458\,b) and DW\,Lyn\,b (HS\,0702+6043\,b). Horizontal 
dashed lines separate the stellar, brown dwarf and exoplanet regimes, respectively. 
The cases of zero inclination (face on) and $90$ degrees inclination (edge on) are 
indicated in the figure. Assuming a random distribution of orbital inclinations, the 
arrows at the top show the probabilities $\mathcal P (i < \theta) = 1 - \cos\theta$ that
the orbital inclination $i$ is lower than a certain value $\theta$, e.\ g.\ an
inclination lower than $60$ degrees has a probability of $50$\%. The brown dwarf
candidate V1636\,Ori\,b would be stellar for inclinations below $23.12$
degrees. The possibility for this being the case is $8.03$\%. Accordingly, the
exoplanet candidate DW\,Lyn\,b would be a brown dwarf for inclinations
below $25.42$ degrees (possibility $9.68$\%) and a star for inclinations below $4$
degrees (possibility $0.24$\%).}
\label{inclinations}
\end{figure}

After subtraction of the evolutionary component we find sinusoidal signatures
in the O--C diagram of both frequencies \smash{$f_1$} and \smash{$f_2$}. The
most plausible explanation for this sinusoidal behavior is again an orbital
origin and a companion candidate around DW\,Lyn. Figure~\ref{hs0702f1}
shows the O--C diagrams for \smash{$f_1$} and \smash{$f_2$}. Within the error bars, 
the same signature consistent in amplitude, phase and period is also seen in the 
O--C diagram for the independent frequency $f_2$. Weighting over the amplitudes and 
periods in the O--C diagrams of $f_1$ and $f_2$ results in a minimum companion
mass of $5.58\pm 1.44$ Jupiter masses for this candidate and an orbital
separation of $1.148\pm 0.050$ AU. Hence, DW\,Lyn\,b would most likely be
in the exoplanet mass regime (see Fig.~\ref{inclinations} for the effect of
the unknown inclination on the true mass).
\section{Summary}
We present first results for the two EXOTIME targets
V1636\,Ori and DW\,Lyn, assuming that the parabolic
component in the O--C diagram is due to evolution and 
that the sinusoidal residuals are the signature of 
sub-stellar companion candidates.
The secular period change in the frequencies $f_1$ and $f_2$ in V1636\,Ori
reveals an expansion of the envelope with a timescale consistent with core
helium exhaustion. The same conclusions are drawn for DW\,Lyn, also based
on the two strongest frequencies $f_1$ and $f_2$ in this star.
Concerning the search for sub-stellar companions, in both targets we find evidence
for the presence of a companion in a rather wide orbit. The brown dwarf candidate
V1636\,Ori\,b has a minimum mass of $31.41$ Jupiter masses and orbits the host star
at a separation of $0.269$ AU. The candidate DW\,Lyn\,b is most likely an exoplanet
with a minimum mass of $5.58$ Jupiter masses and with an orbital separation of
$1.148$ AU.    
\acknowledgements
The authors thank all the observers in the
\mbox{EXOTIME} collaboration for their
contributions.
RL acknowledges support by MPS IMPRS for this project.
This presentation has benefitted from a 
grant awarded to SS by the conference organisers through their
sponsors ESF (European Science Foundation), HELAS (The
European Helio- and Asteroseismology Network) and EAST (The
European Association for Solar Telescopes).
%

%

\begin{thebibliography}{19}
%
\bibitem[{{Benatti} {et~al.}(2010){Benatti}, {Silvotti}, {Claudi}, {Schuh},
  {Lutz}, {Kim}, {Janulis}, {Papar{\`o}}, {Baran}, \&
  {{\O}stensen}}]{2010arXiv1012.0747B}
{Benatti}, S., {Silvotti}, R., {Claudi}, R.~U., {et~al.} 2010, ArXiv e-prints,
  arXiv:1012.0747
%
\bibitem[{{Beuermann} {et~al.}(2012{\natexlab{a}}){Beuermann}, {Breitenstein},
  {Bski}, {Diese}, {Dubovsky}, {Dreizler}, {Hessman}, {Hornoch}, {Husser},
  {Pojmanski}, {Wolf}, {Wo{\'z}niak}, {Zasche}, {Denk}, {Langer}, {Wagner},
  {Wahrenberg}, {Bollmann}, {Habermann}, {Haustovich}, {Lauser}, {Liebing}, \&
  {Niederstadt}}]{2012A&A...540A...8B}
{Beuermann}, K., {Breitenstein}, P., {Bski}, B.~D., {et~al.}
  2012{\natexlab{a}}, \aap, 540, A8
%
\bibitem[{{Beuermann} {et~al.}(2012{\natexlab{b}}){Beuermann}, {Dreizler},
  {Hessman}, \& {Deller}}]{2012A&A...543A.138B}
{Beuermann}, K., {Dreizler}, S., {Hessman}, F.~V., \& {Deller}, J.
  2012{\natexlab{b}}, \aap, 543, A138
%
\bibitem[{{Charpinet} {et~al.}(2011){Charpinet}, {Fontaine}, {Brassard},
  {Green}, {van Grootel}, {Randall}, {Silvotti}, {Baran}, {{\O}stensen},
  {Kawaler}, \& {Telting}}]{2011Natur.480..496C}
{Charpinet}, S., {Fontaine}, G., {Brassard}, P., {et~al.} 2011, \nat, 480, 496
%
\bibitem[{{Geier} {et~al.}(2012){Geier}, {Classen}, {Br{\"u}nner}, {Nagel},
  {Schaffenroth}, {Heuser}, {Heber}, {Drechsel}, {Edelmann}, {Koen}, {O'Toole},
  \& {Morales-Rueda}}]{2012ASPC..452..153G}
{Geier}, S., {Classen}, L., {Br{\"u}nner}, P., {et~al.} 2012, in 
  ASP Conference Series, 
  Vol. 452, Fifth Meeting on Hot
  Subdwarf Stars and Related Objects, eds. D.~{Kilkenny}, C.~S. {Jeffery}, \&
  C.~{Koen}, 153
%
\bibitem[{{Han} {et~al.}(2003){Han}, {Podsiadlowski}, {Maxted}, \&
  {Marsh}}]{2003MNRAS.341..669H}
{Han}, Z., {Podsiadlowski}, P., {Maxted}, P.~F.~L., \& {Marsh}, T.~R. 2003,
  \mnras, 341, 669
%
\bibitem[{{Han} {et~al.}(2002){Han}, {Podsiadlowski}, {Maxted}, {Marsh}, \&
  {Ivanova}}]{2002MNRAS.336..449H}
{Han}, Z., {Podsiadlowski}, P., {Maxted}, P.~F.~L., {Marsh}, T.~R., \&
  {Ivanova}, N. 2002, \mnras, 336, 449
%
\bibitem[{{Heber}(2009)}]{2009ARA&A..47..211H}
{Heber}, U. 2009, \araa, 47, 211
%
\bibitem[{{Kawaler}(2010)}]{2010AN....331.1020K}
{Kawaler}, S.~D. 2010, AN, 331, 1020
%
\bibitem[{{Lee} {et~al.}(2009){Lee}, {Kim}, {Kim}, {Koch}, {Lee}, {Kim}, \&
  {Park}}]{2009AJ....137.3181L}
{Lee}, J.~W., {Kim}, S.-L., {Kim}, C.-H., {et~al.} 2009, \aj, 137, 3181
%
\bibitem[{{Lutz}(2011)}]{2011PhDT........85L}
{Lutz}, R. 2011, PhD thesis, Georg-August-Universit{\"a}t G{\"o}ttingen
%
\bibitem[{{Lutz} {et~al.}(2011){Lutz}, {Schuh}, \&
  {Silvotti}}]{2011AIPC.1331..155L}
{Lutz}, R., {Schuh}, S., \& {Silvotti}, R. 2011, in 
  AIP Conference Series, 
  Vol. 1331, Planetary Systems beyond the Main
  Sequence, eds. S.~{Schuh}, H.~{Drechsel}, \& U.~{Heber}, 155
%
\bibitem[{{Nordhaus} {et~al.}(2010){Nordhaus}, {Spiegel}, {Ibgui}, {Goodman},
  \& {Burrows}}]{2010MNRAS.408..631N}
{Nordhaus}, J., {Spiegel}, D.~S., {Ibgui}, L., {Goodman}, J., \& {Burrows}, A.
  2010, \mnras, 408, 631
%
\bibitem[{{Qian} {et~al.}(2012){Qian}, {Zhu}, {Dai}, {Fern{\'a}ndez-Laj{\'u}s},
  {Xiang}, \& {He}}]{2012ApJ...745L..23Q}
{Qian}, S.-B., {Zhu}, L.-Y., {Dai}, Z.-B., {et~al.} 2012, \apjl, 745, L23
%
\bibitem[{{Qian} {et~al.}(2009){Qian}, {Zhu}, {Zola}, {Liao}, {Liu}, {Li},
  {Winiarski}, {Kuligowska}, \& {Kreiner}}]{2009ApJ...695L.163Q}
{Qian}, S.-B., {Zhu}, L.-Y., {Zola}, S., {et~al.} 2009, \apjl, 695, L163
%
\bibitem[{{Schuh} {et~al.}(2010){Schuh}, {Silvotti}, {Lutz}, {Loeptien},
  {Green}, {{\O}stensen}, {Leccia}, {Kim}, {Fontaine}, {Charpinet},
  {Franc{\oe}ur}, {Randall}, {Rodr{\'{\i}}guez-L{\'o}pez}, {van Grootel},
  {Odell}, {Papar{\'o}}, {Bogn{\'a}r}, {P{\'a}pics}, {Nagel}, {Beeck},
  {Hundertmark}, {Stahn}, {Dreizler}, {Hessman}, {Dall'Ora}, {Mancini},
  {Cortecchia}, {Benatti}, {Claudi}, \& {Janulis}}]{2010Ap&SS.329..231S}
{Schuh}, S., {Silvotti}, R., {Lutz}, R., {et~al.} 2010, \apss, 329, 231
%
\bibitem[{{Silvotti} {et~al.}(2010){Silvotti}, {Randall}, {Dhillon}, {Marsh},
  {Savoury}, {Schuh}, {Fontaine}, \& {Brassard}}]{2010AN....331.1034S}
{Silvotti}, R., {Randall}, S.~K., {Dhillon}, V.~S., {et~al.} 2010,
  AN, 331, 1034
%
\bibitem[{{Silvotti} {et~al.}(2007){Silvotti}, {Schuh}, {Janulis}, {Solheim},
  {Bernabei}, {{\O}stensen}, {Oswalt}, {Bruni}, {Gualandi}, {Bonanno},
  {Vauclair}, {Reed}, {Chen}, {Leibowitz}, {Paparo}, {Baran}, {Charpinet},
  {Dolez}, {Kawaler}, {Kurtz}, {Moskalik}, {Riddle}, \&
  {Zola}}]{2007Natur.449..189S}
{Silvotti}, R., {Schuh}, S., {Janulis}, R., {et~al.} 2007, \nat, 449, 189
%
\bibitem[{{Soker}(1998)}]{1998AJ....116.1308S}
{Soker}, N. 1998, \aj, 116, 1308
%
\end{thebibliography}
%
\end{document}